# Pressure-tunable magnetic topological phases in magnetic topological insulator MnSb$_4$Te$_7$

Lingling Gao[1#], Juefei Wu[1#], Ming Xi[2#], Cuiying Pei[1], Qi Wang[1,3], Yi Zhao[1], Shangjie Tian[2], Changhua Li[1], Weizheng Cao[1], Yulin Chen[1,3,4], Hechang Lei[2*], and Yanpeng Qi[1,3,5*]

1. School of Physical Science and Technology, ShanghaiTech University, Shanghai 201210, China
2. Department of Physics and Beijing Key Laboratory of Opto-electronic Functional Materials & Micro-nano Devices, Renmin University of China, Beijing 100872, China
3. ShanghaiTech Laboratory for Topological Physics, ShanghaiTech University, Shanghai 201210, China
4. Department of Physics, Clarendon Laboratory, University of Oxford, Parks Road, Oxford OX1 3PU, UK
5. Shanghai Key Laboratory of High-resolution Electron Microscopy, ShanghaiTech University, Shanghai 201210, China

# These authors contributed to this work equally.
* Correspondence should be addressed to Y.P.Q. (qiyp@shanghaitech.edu.cn) or H.C.L. (hlei@ruc.edu.cn)

**ABSTRACT**

**Magnetic topological insulators, possessing both magnetic order and topological electronic structure, provides an excellent platform to research unusual physical properties. Here, we report a high-pressure study on the anomalous Hall effect of magnetic TI MnSb$_4$Te$_7$ through transports measurements combined with first-principle theoretical calculations. We discover that the ground state of MnSb$_4$Te$_7$ experiences a magnetic phase transition from the A-type antiferromagnetic state to ferromagnetic dominating state at 3.78 GPa, although its crystal sustains a rhombohedral phase under high pressures up to 8 GPa. The anomalous Hall conductance $\sigma_{xy}^A$ keeps around 10 $\Omega^{-1}$ cm$^{-1}$, dominated by the intrinsic mechanism even after the magnetic phase transition. The results shed light on the intriguing magnetism in MnSb$_4$Te$_7$ and pave the way for further studies of the relationship between topology and magnetism in topological materials.**

Quantum anomalous Hall effect (QAHE) is induced by the spontaneous magnetization and considered as the quantized version of the conventional anomalous Hall effect (AHE), which have potential applications to the next-generation low consumption devices. The studies of QAHE are important for both fundamental research and industrial applications. Theoretical works predicted that the QAHE could exist in time-reversal symmetry-broken magnetic topological insulators (TIs)[1-3]. The first experimental evidence of QAHE is in Cr-doped (Bi, Sb)$_2$Te$_3$ thin films, grown by molecular beam epitaxy (MBE)[4, 5]. However, the inhomogeneous magnetic dopants could also act as impurities[6] and limit the quality of the magnetic TIs, causing a challenge in the precise control of the multiple elements ratio. Hence, the intrinsic magnetic TIs, where the magnetic atoms are on the lattice sites, are desired for QAHE studies.

Recently, natural layered magnetic TIs of (MnBi$_2$Te$_4$)$_m$(Bi$_2$Te$_3$)$_n$ have attracted extensive attention due to the diverse magnetic topological phases[7, 8]. The odd-layered MnBi$_2$Te$_4$ thin flakes[9] and MnBi$_2$Te$_4$/Bi$_2$Te$_3$ superlattice[10] are the intrinsic magnetic TIs to realize QAHE. Besides, the Sb-doped Mn(Bi$_{0.7}$Sb$_{0.3}$)$_4$Te$_7$ single crystal is a magnetic TI with the ferromagnetic (FM) ground state, which can align the opposite magnetic moments without the magnetic field[11, 12]. Upon compression, the ambient antiferromagnetic (AFM) state of MnBi$_2$Te$_4$ and MnBi$_4$Te$_7$ are suppressed[13-15], while the MnBi$_6$Te$_{10}$ experiences a AFM state to FM state transition, where the intralayer exchange coupling plays an essential role[15].

(MnSb$_2$Te$_4$)$_m$(Sb$_2$Te$_3$)$_n$, the sister system of (MnBi$_2$Te$_4$)$_m$(Bi$_2$Te$_3$)$_n$, also exhibits a versatile magnetism and band topology. The AHE was successfully observed in mechanical exfoliated MnSb$_2$Te$_4$ (m = 1, n = 0) thin flakes[16, 17] and the MnSb$_4$Te$_7$ (m = 1, n = 1) is an AFM topological insulator that can be regulated to FM Weyl semimetal by carrier doping or magnetic field[18]. Moreover, Yin et al.[19] observed an insulator-metal transition at about 9 GPa in MnSb$_2$Te$_4$ (m = 1, n = 0). Interestingly, our recent study shows that there is a superconducting transition at around 30 GPa in MnSb$_4$Te$_7$ (m = 1, n = 1)[20]. Comparing with (MnBi$_2$Te$_4$)$_m$(Bi$_2$Te$_3$)$_n$[14, 15], the (MnSb$_2$Te$_4$)$_m$(Sb$_2$Te$_3$)$_n$ family remains much less explored. Specifically, whether the pressure could induce magnetic structure transition remains unknown, thus it is intriguing to study the magnetic structure and the roles of disparate magnetic coupling in (MnSb$_2$Te$_4$)$_m$(Sb$_2$Te$_3$)$_n$ family under high pressure.

In this work, we study the pressure effect on the magnetism and AHE of MnSb$_4$Te$_7$

combining the electrical and Hall transports measurements with the first-principles calculations. Both the butterfly-shaped magnetoresistance (MR) and the hysteresis loops of Hall resistivity $\rho_{yx}^A(\mu_0 H)$ illustrate the pressure-induced AFM-FM transition in MnSb$_4$Te$_7$ at 3.78 GPa, which further verified by the first-principles calculations. Meanwhile, the critical transition temperatures of AFM-paramagnetic (PM) state or FM-PM state are insensitive to the pressure (~ 14.5 K within 8 GPa). Moreover, the anomalous Hall conductance (AHC) $\sigma_{xy}^A$ keeps around 10 $\Omega^{-1}$ cm$^{-1}$, indicated that intrinsic mechanism of AHE is still dominant even after the magnetic transition.

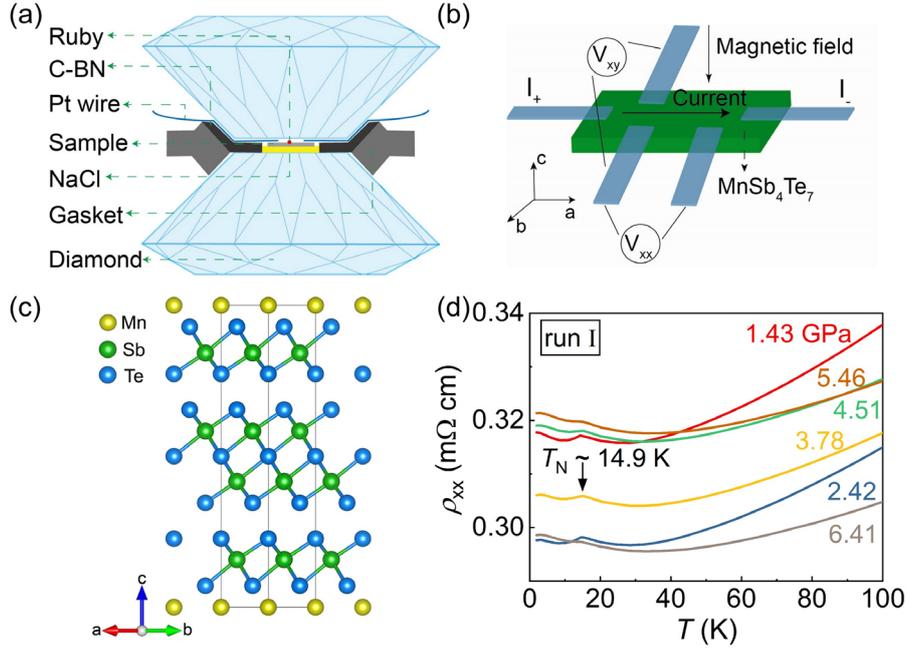

FIG. 1. (a) Schematic of the high-pressure diamond anvil cell (DAC) used for electrical measurements. (b) Schematic of the five-probe measurements configuration in the DAC. (c) Crystal structure of MnSb$_4$Te$_7$. (d) Temperature dependence of electric resistivity $\rho_{xx}$ (T) under high pressures in run I.

Single crystals of MnSb$_4$Te$_7$ were grown via self-flux method and the details of crystal growth are illustrated in Ref. 20[20]. The high-pressure electrical transport measurements were carried on Quantum Design PPMS-9T. High-pressure transport measurements were performed in a nonmagnetic diamond anvil cell[21, 22] (DAC). The schematic plot of DAC electrical transport measurement device is displayed in Figs. 1(a) and (b). A cubic BN/epoxy mixture layer was inserted between BeCu gaskets and electrical leads as insulator layer. A freshly cleaved single-crystal piece was loaded with NaCl powder as the pressure transmitting medium. A five-probe method was used to measure the longitudinal and Hall electrical resistivity. The pressure was determined by the ruby luminescence method[23]. In order to separate the longitudinal and Hall resistivity, we process signals using standard symmetrization and anti-symmetrization procedures[9]. As for first-principles calculations, the Vienna *Ab initio* Simulation Package (VASP) employs in the framework of density functional theory (DFT) [24-26] with PBE-GGA functionals[27]. We consider the spin-orbit coupling (SOC), the DFT-D3 (BJ damping) functional[28] and DFT+$U$[29,30] with the $U_{eff}$ = 5.0 eV, as in Ref. 15. The kinetic-energy cutoff is 500 eV, the Brillouin zone is sampled with $2\pi \times 0.03$ Å$^{-1}$. The total energy converged to $10^{-6}$ eV, and all forces are converged to 0.003 eV/Å.

MnSb$_4$Te$_7$ adopts a rhombohedral structure with space group of $P\bar{3}m1$.[18,20] As shown in Fig. 1(c), the basic unit of MnSb$_4$Te$_7$ are Sb$_2$Te$_3$ quintuple layers (QLs) and Te-Bi-Te-Mn-Te-Bi-Te septuple layers (SLs), which alternates the stack along the *c*-axis direction. The Mn$^{2+}$ ions are FM coupling in each SL with an out-of-plane easy axis, while the adjacent SLs are AFM coupling, establishing an A-type AFM state. Fig. 1(d) and Fig. S1 (supplementary material) display the temperature dependence of longitudinal resistivity $\rho_{xx}(T)$ under various pressures. The Néel temperature ($T_N$) of MnSb$_4$Te$_7$ is at 14.8 K at ambient pressure, which corresponds to the kink in the longitudinal resistivity $\rho_{xx}(T)$ curves. The kink persists at around 14.5 K within the investigated pressure range. The robust of kink in MnSb$_4$Te$_7$ is notably different from MnBi$_4$Te$_7$, which gradually suppressed by pressure.[13,15] The kink in MnSb$_4$Te$_7$ gradually fades with increasing pressure and disappears completely until 8 GPa. The results shown here are consistent with our previous ones using the van der Pauw method[20].

Then, we measure the magneto-transport properties at various pressures, the detailed results of run I are shown in Fig. S2 (supplementary material). As plotted in Figs. 2(a) and (c), we normalize the MR data with the expression MR = [$\rho_{xx}(\mu_0 H) - \rho_{xx}(0)$] / $\rho_{xx}(0)$ × 100%. The negative value of MR is due to the suppression of spin scattering via a magnetic field[31]. At 0.27 GPa, we can observe the resistivity plateau at 2 K as shown inset of Fig. 2(a), which suggests the AFM state at lower pressure region. The abrupt drop of MR is around 0.22 T below the transition temperature, indicating a magnetic field induced AFM-FM transition. It should be noted that the MR results of MnSb$_4$Te$_7$ at 0.27 GPa have analogous characteristics as the ambient results in Ref. 18. Nevertheless, the abrupt drop feature of MR curves evolves into the smooth curves at 4.51 GPa [Fig. 2(c)], suggesting that the FM state may become dominant under high pressure. According to Fig. S2 (supplementary material), the transition pressure is at about 3.78 GPa.

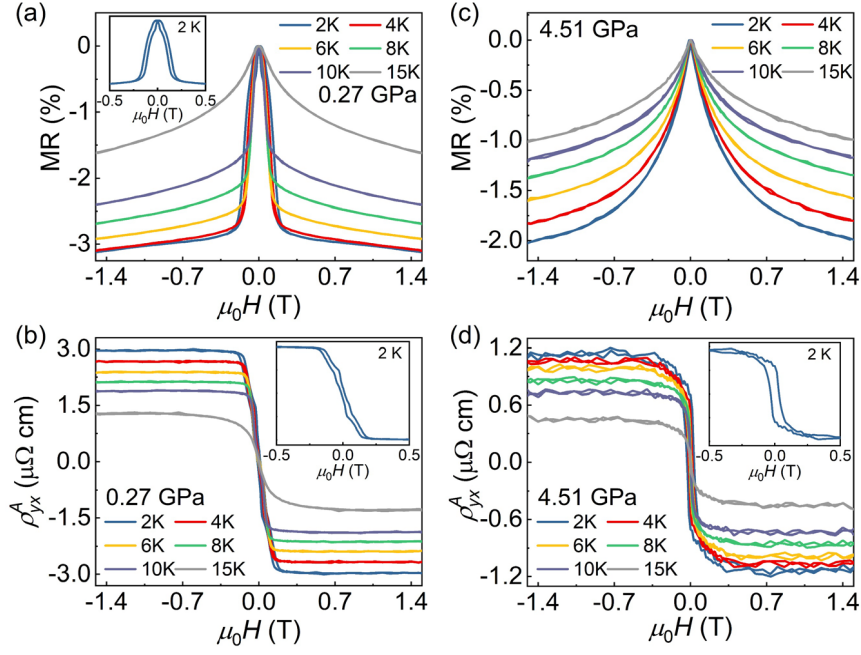

FIG. 2. (a) and (c) Field dependence of magneto-resistance (MR) at various temperature for 0.27 and 4.51 GPa, respectively. The inset of (a) shows MR at 2 K for 0.27 GPa. (b) and (d) The curves of anomalous Hall resistivity $\rho_{yx}^A(\mu_0 H)$ at various temperature for 0.27 and 4.51 GPa, respectively. The inset of (b) and (d) shows $\rho_{yx}^A(\mu_0 H)$ at 2 K for 0.27 GPa and 4.51 GPa, respectively.

The detailed Hall resistivity data at various pressures for run I are shown in Fig. S3 (supplementary material). The Hall resistivity $\rho_{yx}(\mu_0 H)$ includes two parts[32, 33],

$$\rho_{yx} = \rho_{yx}^O + \rho_{yx}^A, \qquad (1)$$

$\rho_{yx}^O = R_H B$ is the ordinary Hall resistivity and $R_H$ is the Hall coefficient, $\rho_{yx}^A$ is the anomalous Hall resistivity. To obtain $\rho_{yx}^A(\mu_0 H)$, we fit the linear component of the Hall resistivity $\rho_{yx}(\mu_0 H)$ under high field and subtract the linear ordinary Hall resistivity[34-36], as plotted in Fig. S4 (supplementary material). The MnSb$_4$Te$_7$ always exhibits AHE within the measured pressure range. At 0.27 GPa [Fig. 2(b)], the saturation field is ~ 0.22 T. The hysteresis loops of $\rho_{yx}^A(\mu_0 H)$ at 2 K have two kinks about 0.03 T and the magnitude gradually diminish with the increase of temperature. This is consistent with the ambient results[18]. Upon further increasing pressure to 4.51 GPa, two kinks of the hysteresis loops of $\rho_{yx}^A(\mu_0 H)$ at 2K vanish [inset of Fig. 2(d)] and the hysteresis loops exhibit the typical FM behaviors. In a word, the high-pressure measurements of MR and $\rho_{yx}^A(\mu_0 H)$ suggest a pressure-induced transition from AFM state to FM dominating state.

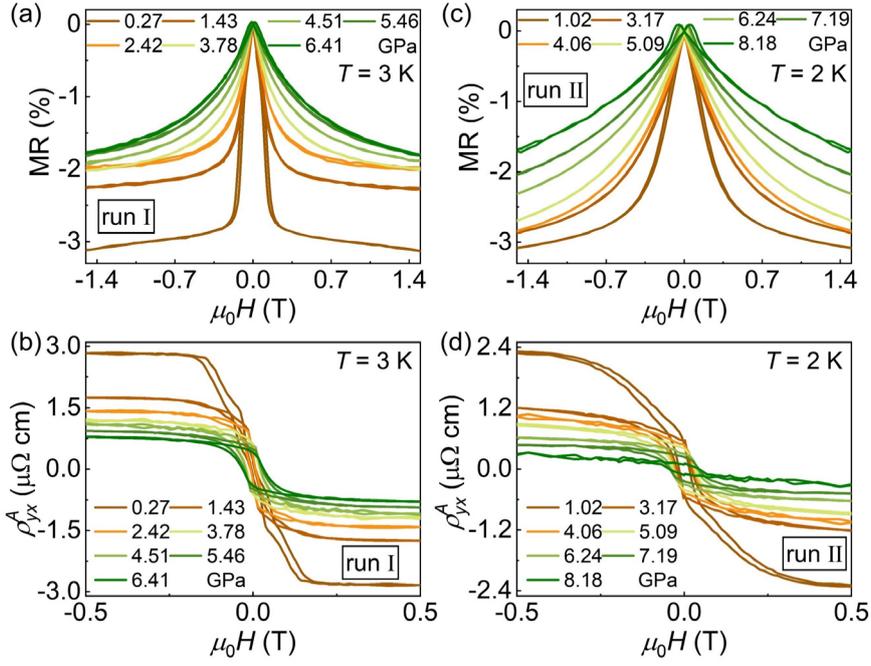

FIG. 3. (a) Field dependence of MR at various pressures in run I. (b) The curves of $\rho_{yx}^A$ ($\mu_0H$) at various pressures in run I. (c) Field dependence of MR at various pressures in run II. (d) The curves of $\rho_{yx}^A$ ($\mu_0H$) at various pressures in run II.

To further illustrate the features of this AFM-FM transition, we summarize the results of MR and $\rho_{yx}^A$($\mu_0H$) at various pressures, as shown in Fig. 3. The magnitude of MR gradually decreases with increasing pressure in different runs [Figs. 3(a) and (c)]. We can observe that the abrupt drop in MR curves fades away around 3.78 GPa [Fig. 3(a)] and MR curves persist butterfly-shape above 4.06 GPa [Fig. 3(d)], strongly indicate the appearance of FM state at high-pressure region. As for $\rho_{yx}^A$($\mu_0H$), it decreases from ~2.4 μΩ cm at 1.02 GPa to ~ 0.28 μΩ cm at 8.18 GPa, while the hysteresis loops of $\rho_{yx}^A$($\mu_0H$) are widened above 3.78 GPa [Figs. 3(b) and (d)]. The hysteresis loops persist FM characteristics above 3.78 GPa in both run I and II, indicating the formation of FM dominating state[37, 38]. This is resemble to the magnetic phase transition in $MnBi_6Te_{10}$ under high pressure[15].

To better understand the magnetic phase transition, we perform first-principles calculations to give a qualitative description. As plotted in Fig. 4(a), the enthalpy of FM state is smaller than that of AFM state around 12 GPa, suggesting that the FM interaction is dominant under high-pressure range. Considering that the intralayer interaction is FM in both FM and AFM $MnSb_4Te_7$ [the inset of Fig. 3(a)], the energy difference reflects the competition of FM and AFM coupling between interlayers. Since the interlayer coupling is on the order of $10^{-1}$ meV/atom within 20 GPa, DFT calculations may give higher transition of pressure than that in experiment. Besides, we recheck our calculations by adjusting vdW effect and $U_{\text{eff}}$, and observe similar tendency shown in Fig. 3(a). Thus, the theory confirms that pressure could induce the magnetic transition in principle.

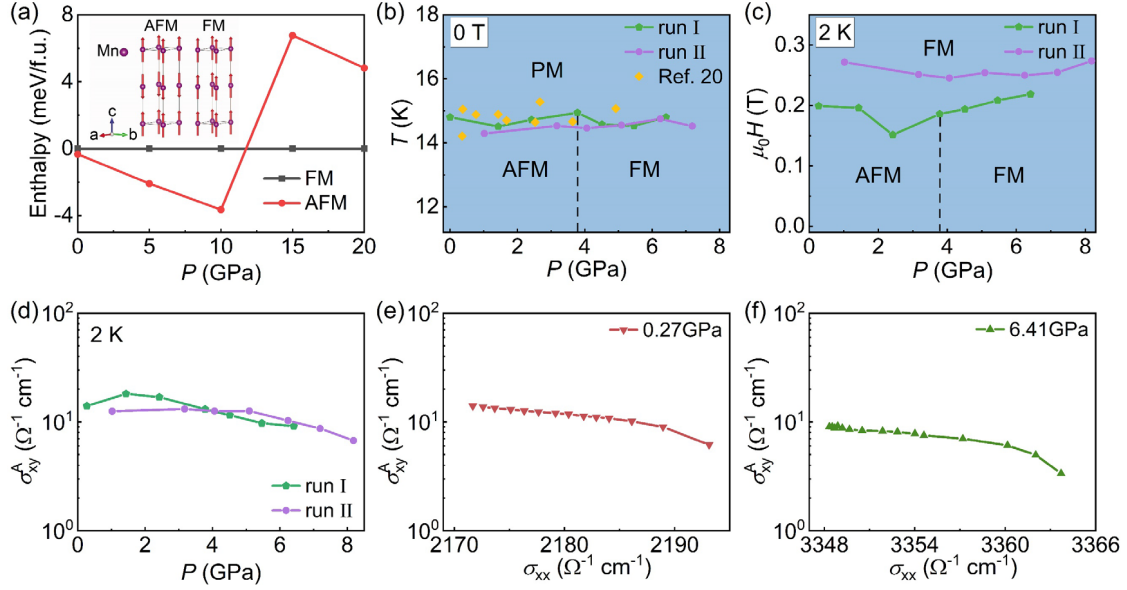

FIG. 4. (a) Calculated enthalpy of the AFM phase relative to that of the FM phase at pressures. (b) The temperature-pressure phase diagram at 0 T. (c) The magnetic field-pressure phase diagram at 2 K. PM, AFM and FM represents for paramagnetic, antiferromagnetic and ferromagnetic phase, respectively. (d) Pressure-dependent anomalous Hall conductivity $\sigma_{xy}^A$ in different runs. $\sigma_{xy}^A$ as a function of the longitudinal conductivity $\sigma_{xx}$ at 0.27 GPa (e) and 6.41 GPa (f).

Fig. 4(b) plots the pressure dependence of magnetic transition temperature determined from different runs. Transition temperature keeps around 14.5 K within the investigated pressure range and becomes indistinguishable above 8 GPa. The saturation magnetic field $H_S$ as a function of pressure at 2 K are also summarized in Fig. 4(c), which are extracted from $\rho_{yx}(\mu_0 H)$ curves in Fig. S3 (supplementary material). When the ground state of MnSb$_4$Te$_7$ is AFM, the $H_S$ decreases with the increase of pressure from 0.2 T at 0.27 GPa to 0.15 T at 2.42 GPa in run I. The change of $H_S$ is inconsistent with the excepted enhanced AFM interlay coupling with compression along the $c$-axis[15]. In the FM dominating state, the $H_S$ begins to increase with the pressure. The $H_S$ reaches about 0.27 T at 8.18 GPa, indicating pressure-induced enhancement of FM exchange coupling. Fig. 4(d) displays the AHC of the magnetic field induced FM Weyl semimetal state[18] as a function of pressure at 2 K. The $\sigma_{xy}^A$ [$= \rho_{xy}^A / (\rho_{xy}^{A2} + \rho_{xx}^2)$] shows a slow decrease with pressure, and keeps about 10 $\Omega^{-1}$ cm$^{-1}$, which is the same order of magnitude as ambient pressure. Moreover, the $\sigma_{xy}^A$ is overall independent of longitudinal conductivity $\sigma_{xx}$ under different pressures, as shown in Figs. 4(e) and 4(f). Our results suggest that the intrinsic mechanism still dominates the AHE at high pressure[31, 39, 40].

In conclusion, pressure could modulate the magnetism of topological insulator candidate MnSb$_4$Te$_7$. The butterfly-shaped MR and the hysteresis loops of $\rho_{yx}^A$ ($\mu_0H$) confirms that AFM ground state evolves to the FM dominating state at 3.78 GPa, and the first-principles calculations qualitatively support these experimental observations. The AHC keeps about 10 $\Omega^{-1}$ cm$^{-1}$, which still dominates by the intrinsic mechanism under high pressure. Our results shed light on the possibility to modulate the topology and magnetism in the magnetic topological insulator by pressure, and provide a platform for further understanding the relation between the topological phase and magnetic order.

See the supplementary material for detailed data of electric resistivity as a function of temperature in run II, and longitudinal resistivity, Hall resistivity and anomalous Hall resistivity as a function of magnetic field at different pressures and temperatures in run I.

This work was supported by the National Natural Science Foundation of China (Grant No. 12004252, U1932217, 52272265, 11974246), the National Key R&D Program of China (Grant No. 2018YFA0704300 and 2022YFA1403800), and Shanghai Science and Technology Plan (Grant No. 21DZ2260400). The authors thank the support from Analytical Instrumentation Center (# SPST-AIC10112914), SPST, ShanghaiTech University. The calculations were carried out at the HPC Platform of ShanghaiTech University Library and Information Services.

## AUTHOR DECLARATIONS

### Conflict of Interest

The authors have no conflicts to disclose.

### DATA AVAILABILITY

The data that support the findings of this study are available from the corresponding authors upon reasonable request.